\documentclass[twocolumn,showpacs,amssymb,aps,prl]{revtex4}
\usepackage{graphicx}
\usepackage{dcolumn}
\usepackage{bm}

\renewcommand{\vec}[1]{\bm{#1}}
\newcommand{\tens}[1]{\mbox{\textbf{\textit{\textsf{#1}}}}}

\newcommand{\mi}{\mathrm{i}}

\begin{document}

\title{Probing atom-surface interactions by diffraction of Bose-Einstein
condensates}

\author{Helmar Bender}
\author{Christian Stehle}
\author{Claus Zimmermann}
\author{Sebastian Slama}
\email{slama@pit.physik.uni-tuebingen.de}
\affiliation{Physikalisches Institut and Center for Collective Quantum Phenomena in LISA+, Universit\"at
T\"ubingen, Auf der Morgenstelle 14, D-72076 T\"ubingen, Germany}
\author{Johannes Fiedler}
\author{Stefan Scheel}
\affiliation{Institut f\"ur Physik, Universit\"at Rostock, Universit\"atsplatz
3, D-18055 Rostock, Germany}
\author{Stefan Yoshi Buhmann}
\email{s.buhmann@imperial.ac.uk}
\affiliation{Quantum Optics and Laser Science, Blackett Laboratory,
Imperial College London, Prince Consort Road, London SW7 2AZ, United Kingdom}

\author{Valery N. Marachevsky}
\email{maraval@mail.ru}
\affiliation{Institute of Theoretical and Mathematical Physics, Saint-Petersburg State University,
198504 St.Petersburg,  Russia}

\date{\today}

\begin{abstract}
In this article we analyze the Casimir--Polder interaction of atoms with a solid
grating and an additional repulsive interaction between the atoms and the
grating in the presence of an external laser source. The combined potential
landscape above the solid body is probed locally by diffraction of Bose-Einstein
condensates. Measured diffraction efficiencies reveal information about the
shape of the Casimir--Polder interaction and allow us to discern between models
based on a pairwise-summation (Hamaker) approach and Lifshitz theory.
\end{abstract}


\maketitle 

The Casimir--Polder (CP) interaction is one of a class of examples where
fluctuating electromagnetic fields give rise to (normally
attractive) forces between matter \cite{LennardJones32,Casimir48}. For
infinitely extended plane surfaces CP forces can be readily calculated from the
polarizability of the atom and the dielectric properties of the substrate
\cite{Scheel08} and have been measured in a number of experiments
\cite{Sukenik93,Shimizu01,Druzhinina03,Pasquini04,Obrecht07,Bender10}.
However, of particular importance is the influence of the surface geometry
\cite{Rodriguez11,Marachevsky12}. Non-trivial geometries can have a large
impact on the exact force profile and can potentially be used for manipulating
the closely related Casimir forces \cite{Levin10}.
The possibility to tailor the Casimir force is also of importance for
applications in the MEMS and NEMS industry where it is one of the limiting
factors in the miniaturization of micromachines and microsensors \cite{Rio05}.

\begin{figure}[h]
\centerline{\includegraphics[width=\columnwidth]{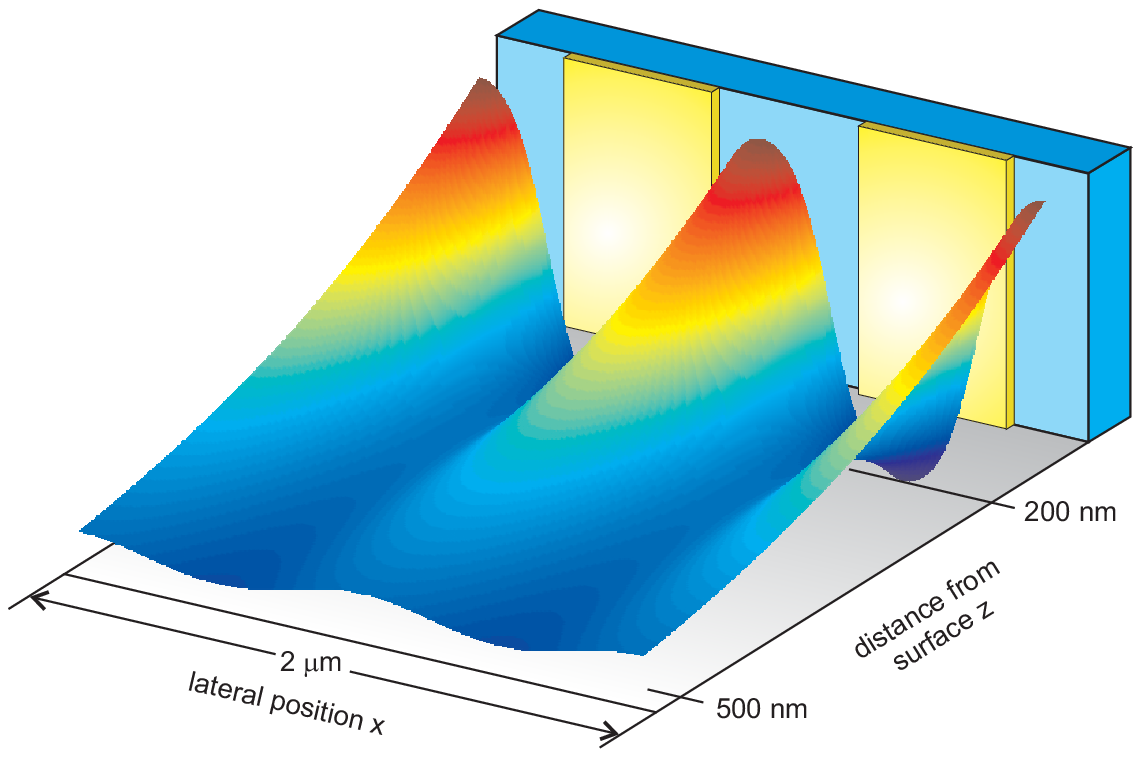}}
\caption{Sketch of the metal grating including the combined
Casimir-Polder and evanescent wave potential landscape as calculated from
Eqs.~(\ref{CPpotential}) and (\ref{Uevanescent1}) within Lifshitz theory for a
laser power of $P=200~\textrm{mW}$. The distances from the surface range from
$z=200~\textrm{nm}$ to $z=500~\textrm{nm}$, the lateral position span is
$2~\mu\textrm{m}$ and the potential modulation at a distance of
$z=200~\textrm{nm}$ is $\Delta E/k_\mathrm{B}=14~\mu\textrm{K}$.}
\label{fig1}
\end{figure}

One class of non-trivial geometries that have been investigated theoretically
both in the framework of atom-surface and surface-surface interactions are
periodic structures such as gratings \cite{Marachevsky08,Reyes10,Davids10}.
Experimentally, CP forces above gratings were measured by different
methods \cite{Pasquini06,Zhao08,Grisenti99,Oberst05,Perreault05}. In those
experiments, the power law coefficients describing the CP potential in
the electrostatic and in the retarded regimes were determined. 
However, the measured values represent only an average over the complicated
potential landscape above the structures. In this work we fully account for the
dependence of the potential on the lateral position above such a surface.

Dispersion potentials in nontrivial geometries can be calculated using
the Hamaker approach \cite{deBoer36,Hamaker37}, which is based on a pairwise
summation of van der Waals forces between volume elements of one body with those of
the other. However, such an approach neglects many-body interactions which can
lead to wrong results in particular for complex geometric
structures \cite{Thennadil01,Marachevsky09}. The non-additivity of Casimir
forces induced by many-body interactions \cite{Buhmann06} implies that the position-, shape- and
material-dependencies of such forces are intertwined in a complicated way.
Nevertheless, the Hamaker approach is widely used in applications such as
colloid science and biology \cite{Parsegian05}. The most prominent example is
the adhesive force of gecko feet \cite{Autumn00}. Recently, experiments have
shown deviations from Hamaker theory in surface-surface interactions
\cite{Marachevsky09}. Surprisingly, not a single experiment has so far addressed
the accuracy of the Hamaker approach in atom-surface interactions.
 
In this Letter we present both simulations and measurements of the potential
landscape for a single atom that is positioned at a submicron distance from a
grating of metal nanowires (see Fig.~\ref{fig1}). Our measurements allow us to
distinguish between results obtained using the Hamaker approach and those using
exact Lifshitz theory.

The potential landscape is composed of an attractive contribution due to the
Casimir--Polder force and a repulsive contribution due to an evanescent light
wave (EW) at the surface. The evanescent wave is generated by internal total
reflection of a laser beam in the dielectric substrate carrying the grating. A
repulsion from the surface is achieved by a laser
($\lambda=765~\textrm{nm}$) that is blue-detuned with respect
to the transition frequency of the atoms (Rb: $\lambda_0=780~\textrm{nm}$).
Recently, we used this setup and enhanced the
evanescent waves by exciting surface plasmon polaritons at the surface
\cite{Stehle11}. Here, we exploit the fact that the exact shape of the total
potential landscape can be tuned by the strength of the optical dipole
potential via the laser intensity. This allows us to acquire spatially
resolved information on the potential landscape.

Figure~\ref{fig1} shows the calculated potential landscape in Lifshitz
theory for a typical laser power of $P=200~\textrm{mW}$ including the optical
dipole potential of the evanescent wave. In the simulations, the ground-state CP
potential of the atoms is calculated as
\cite{Buhmann04}
\begin{equation}
\label{CPpotential}
U_\mathrm{CP}(\vec{r})=\frac{\hbar\mu_0}{2\pi}
 \int_0^\infty d\xi\,\xi^2\alpha(\mi\xi)\,\mathrm{Tr}\,
 \tens{G}^{(1)}(\vec{r},\vec{r},\mi\xi)\;.
\end{equation}
Here, $\alpha(\mi\xi)$ is the isotropic ground-state polarisability of the Rb
atoms and $\tens{G}^{(1)}$ is the scattering Green tensor which, for the grating
structure in Fig.~\ref{fig1}, can be given as a Rayleigh decomposition. 
Due to the integral over all imaginary frequencies as a result of the vacuum
fluctuations of the e.m. field, the CP potential depends on all atomic
transition frequencies and all eigenfrequencies of the macroscopic system
(grating). 

The EW potential
\begin{equation}
\label{Uevanescent1}
U_\mathrm{EW}(\vec{r})
=\sum_{i=1,2}\frac{|\vec{d}_i|^2|\vec{E}(\vec{r})|^2}
 {3\hbar\Delta_i}
\end{equation}
is the potential due to the external monochromatic electric field $\vec{E}$ with
its frequency $\omega$ close 
a specific set of atomic transitions of Rb with dipole matrix elements
$\vec{d}_i$ and detunings $\Delta_i=\omega -\omega_i$. It is dominated by these
atomic transitions and the transmission properties of the grating at a single
laser frequency $\omega$. 

In contrast, the CP potential in the Hamaker approach is calculated in
local-field corrected first-order Born approximation as 
\begin{eqnarray}
\label{UHamaker}
\tens{G}^{(1)}(\vec{\varrho},\omega)&=&\frac{\omega^2}{c^2}
\frac{\chi(\omega)}{1+\chi(\omega)/3} \nonumber\\
&&\times \int \mathrm d^3s\, \tens{R}^{(0)}
(\vec{r},\vec{s},\omega) \tens{R}^{(0)} (\vec{s},\vec{r},\omega)
\end{eqnarray}
where $\tens{R}^{(0)} (\vec{r},\vec{s},\omega)$ is the regular part of
the Green tensor, $\chi(\omega)$ is the susceptibility of the gold stripes, and the
integration extends over the total volume $V$ of the grating. Details of all the
calculations are contained in \cite{Supp}.

%
\begin{figure}[ht]
\centerline{\includegraphics[width=\columnwidth]{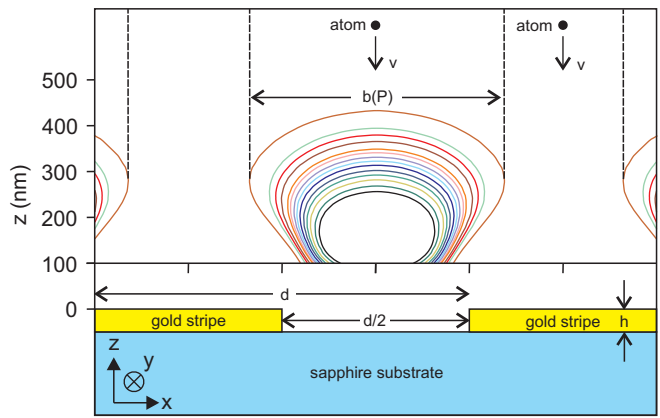}}
\caption{Geometry of the metal grating. Approximately 200 gold stripes
with $h=50~\mathrm{nm}$ height and $500~\mathrm{nm}$ width are deposited on a
sapphire substrate and form a grating with $d=1~\mu\mathrm{m}$ period. 
The combination of the repulsive potential due to the evanescent wave with the
Casimir--Polder interaction forms a potential landscape above the grating.  The
colored lines (black to brown, corresponding to laser powers $P=120$, $126$,
$133$, $138$, $144$, $151$, $156$, $162$, $169$, $174$, $187$, $198$, $211$,
$247~\mathrm{mW}$) are simulations of equipotential lines using Lifshitz theory 
for a $^{87}$Rb atom moving towards the surface with velocity
$v=3.4~\mathrm{cm}/\mathrm{s}$. From those we deduce
the width $b(P)$ where atoms are reflected.}
\label{fig2}
\end{figure}

The CP potential with its strong attraction towards the gold stripes and the
repulsive EW potential with its maximal repulsion above the sapphire surface
combine to the periodic potential landscape in Fig.~\ref{fig1}. It resembles a
chain of hills in front of the grating surface with valleys that lead to the
centers of the gold stripes. The heights and widths of the hills depend on the
laser power (Fig.~\ref{fig2}) with larger powers resulting in
higher and broader hills. Experimentally, we probe the width of the hills by
reflecting Bose--Einstein condensates (BEC) from the surface. 

The experiment is carried out as follows. A BEC is prepared in a magnetic trap
close to the surface of the grating and accelerated such that it moves towards
the surface with a constant velocity $v=3.4~\mathrm{cm}/\mathrm{s}$. The
experimental details of this preparation are contained in \cite{Supp}. The atoms
reflect from the surface only at those lateral positions where the potential
height exceeds the kinetic energy of the atoms. This happens in a zone with
width $b$ in each lattice site (see Fig.~\ref{fig2}). Note that considerable
quantum reflection of Rb atoms at the CP potential of a solid surface would
require atomic velocities below few $\mathrm{mm}/\mathrm{s}$ \cite{Pasquini04}. For the used
velocity of $v=3.4~\mathrm{cm}/\mathrm{s}$ it is completely negligible. By
tuning the laser power, the reflection zone width $b$ is changed and different
distances from the surface are probed. Each atom of the BEC approaching the
surface constitutes a matter wave with a lateral extension that is given by the
size of the BEC on the order of several tens of microns. This size is much
larger than the grating period, thus the matter wave is diffracted from the
periodic structure of reflection zones in a direction $x$ of period $d$. 

In a simplified model that neglects the curvature of the equipotential lines we
consider reflection of the matter wave from the same reflection zones of
width $b$ as for a single atom. The resulting atomic momentum distribution in
the far field is analogous to Fraunhofer diffraction of light and is determined
by the Fourier transform of the step function $\sum_{n} \theta(b/2-|x-nd|)$. In
this density imprinting model, the external potential leads to a reflection of
the matter wave, but does not significantly alter its phase. For a periodic
arrangement of rectangular stripes as shown in Fig.~\ref{fig2}, the reflected
wave is composed of wavevectors $k_x$ with relative occupation $p(k_x)$,
\begin{equation}\label{eq:sinc}
p(k_x)\propto\sum_{n}\delta(k_x-nq) \cdot
 \left|\mathrm{sinc}\left(\frac{\pi k_x}{q}\cdot\frac{b}{d}
 \right)\right|^2~.
\end{equation}

\begin{figure}[ht]
\centerline{\includegraphics[width=\columnwidth]{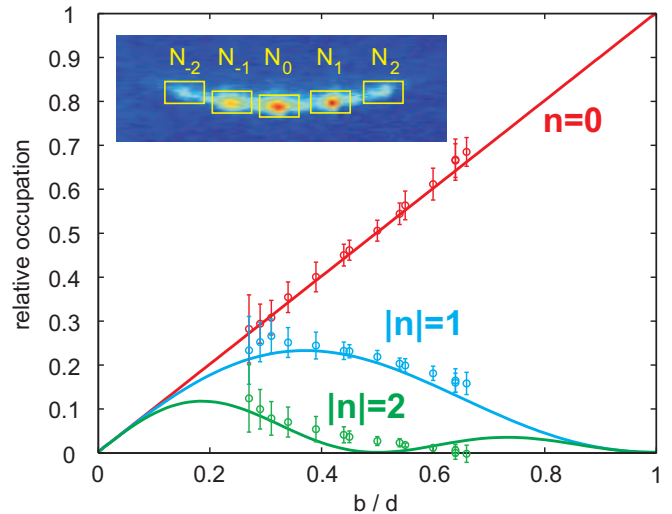}}
\caption{Matter wave diffraction. Relative occupation of diffraction
orders as a function of the relative width of the reflection zone $b/d$. The
curves are obtained from Eq.~(\ref{eq:sinc}). Data points represent the measured
occupation of diffraction orders like shown in the inset. The horizontal
position of each triple of data points (its value $b/d$) is obtained from a fit
to the theoretical curves.}
\label{fig3}
\end{figure}

This expression is a sum over delta functions at integer multiples $n$ of the
lattice vector 
$q=2\pi/d$.
The number $n$ denotes the diffraction order. The sum is multiplied with an
envelope amplitude given by a sinc-function that determines the corresponding
occupation of the diffraction order. The relative occupation of the diffraction
orders depends only on the ratio $b/d$. This is illustrated in the theoretical
curves in Fig.~\ref{fig3}. 
In the limit of $b/d\rightarrow 0$ the situation resembles the emission of
waves from a chain of point-like sources, in which all diffraction orders are
equally occupied. In contrast, the limit $b/d\rightarrow 1$ corresponds to a
reflection from a surface with constant density profile. Here, the atomic cloud
remains fully in the diffraction order $n=0$ with wavevector $k_x=0$.

%

In the experiment, we analyze the relative occupation of individual
diffraction orders by measuring the momentum distribution $p_x=\hbar k_x$ of the
atoms. This is done by taking an absorption image of the cloud after ballistic
expansion for a time-of-flight of $t_\mathrm{tof}=21.5~\mathrm{ms}$ after
reflection from the surface. A typical image is shown in the inset of
Fig.~\ref{fig3}. From the image the atom numbers $N_n$ corresponding to
diffraction orders $n$ are counted within the yellow boxes and are scaled to the
total number of reflected atoms.
This provides us data triples of relative populations of diffraction orders
$n=0,\pm 1,\pm 2$
for each value of laser power. The populations for
$n\neq 0$ are averaged over the populations of the orders with $\pm n$. Each
triple is individually fitted by Eq.~(\ref{eq:sinc}) and is thus attributed a
certain value of $b/d$. The result of the fit is compared with the theoretical
curves in Fig.~\ref{fig3}. The fact that the data points coincide with the
corresponding theory curves for each triple is a signature that the diffraction
process is well described within the simple model.

\begin{figure}[ht]
\centerline{\scalebox{1}{\includegraphics{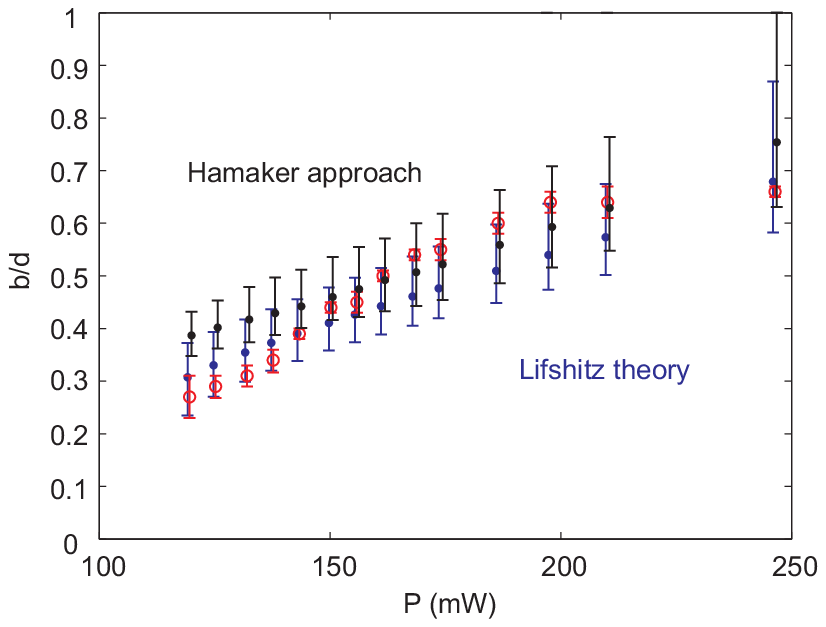}}}
\caption{Comparison between theory and experiment. Width of the
reflection zone $b/d$ versus laser power $P$. Red circles are experimental data
points obtained from the fit in Fig. \ref{fig3}. Error bars of the data points
are due to the combined statistic and systematic uncertainty in the atom number
measurement. Blue dots show the theoretical values taken from the
equipotential lines in Fig.~\ref{fig2} and represent the result of the Lifshitz
model. Black points are the corresponding results based on a Hamaker approach.
Error bars of the theoretical points
represent the systematic error due to the uncertainty in the laser intensity at
the surface $\Delta I=\pm 5\%$ and the velocity of the atoms $\Delta
v=\pm0.3~\mathrm{cm}/\mathrm{s}$. Please note that the data points are
horizontally shifted one line width ($\lesssim 1~$mW) for clarity.}
\label{fig4}
\end{figure}

The fitted values of $b/d$ are now compared with the theoretical prediction
which is accessible from the width of the equipotential lines shown in
Fig.~\ref{fig2}. As can be seen in Fig.~\ref{fig4}, the experimental data agree
with the Lifshitz theory within their statistic and systematic errors. In
contrast, the Hamaker approach underestimates the strength of the
Casimir--Polder potential. The corresponding values of $b/d$ for low values of
$P$ in Fig.~\ref{fig4} are thus larger than the observed data points
and deviate from them by more than one standard deviation. In the range of
large $P$ in Fig.~\ref{fig4} the optical potential dominates over the CP
potential and reduces the difference between Lifshitz theory and Hamaker model.
In this regime the data are compatible with both theoretical models.

The different functional profiles of the line shapes of the data points and the Lifshitz
theory can be attributed to the simplicity of the diffraction model: in particular, the
value of $b/d$ of the measured data points saturates for large laser powers.
This observation is not compatible with the density imprinting model. For high
reflectivities an additional effect comes into play. Here, the assumption of
instantaneuos reflection is not justified. Instead, the interaction time of the
atoms with the surface potential and the strength of the latter depend on the
lateral position $x$, i.e. depending on the lateral position the matter wave
acquires a different phase. A periodic potential imprints a phase that leads to
a substantial diffraction even when all atoms are reflected and thus simulates a
saturation of $b/d$ even for large laser powers \cite{Guenther07,Cronin09}. 

Concluding, we have experimentally probed surface potential landscapes that are
composed of Casimir--Polder forces and optical dipole forces above metallic
nanostructures. We have used matter-wave diffraction of Bose-Einstein
condensates as a measuring tool which, in principle, can be applied to arbitrary
surfaces. Complementary to previous experiments in which spatial averages of the
Casimir--Polder coefficients were determined, we obtain additional spatial
information by analyzing the occupation of individual diffraction orders. Our
data agree quantitatively with numerical calculations of the surface potentials
based on Lifshitz theory, whereas a Hamaker approach leads to incompatible
results for low laser powers.

The fact that we understand these potentials very well is crucial for the design
and realization of nanoscale surface traps for surface quantum optics
experiments with cold atoms. Moreover, the metallic parts of the surface can
give rise to spectrally broad surface plasmon resonances in the optical
frequency range. Related phonon polariton resonances in the infrared frequency
range have e.g. led to the observation of repulsive Casimir--Polder forces of
highly excited Cs atoms \cite{Failache99}. A plasmon-based repulsive
Casimir--Polder force would offer fascinating scenarios for controlling CP
forces \cite{Intravaia07} and for generating surface traps for cold atoms that
do not require external magnetic or optical fields.

C.S. was supported by Carl-Zeiss Stiftung Baden-W\"urttemberg. Support from the
European Science Foundation (ESF) within the activity `New Trends and
Applications of the Casimir Effect' is gratefully acknowledged. This work was
financially supported by the UK EPSRC and
by the COSCALI network within the IRSES program of the European Commission under
Grant No. PIRSES-GA-2010-268717. V.N.M was partially supported by the Saint-Petersburg State University grant No 11.38.660.2013.
Furthermore, we acknowledge support by the Deutsche Forschungsgemeinschaft within 
the European Collaborative Research program of the European Science Foundation.

\end{document}


\title{Probing atom-surface interactions by diffraction of
Bose-Einstein
condensates\\ Supplemental Material}
\maketitle

\section{Simulation of the Casimir-Polder potential}
The ground-state CP potential of an atom can be given as
\cite{Buhmann04}
\begin{equation}
\label{CPpotential2}
U_\mathrm{CP}(\vec{r})=\frac{\hbar\mu_0}{2\pi}
 \int_0^\infty d\xi\,\xi^2\mathrm{Tr}[\bm{\alpha}(\mi\xi)\sprod
 \tens{G}^{(1)}(\vec{r},\vec{r},\mi\xi)]\;.\end{equation}
Here, $\bm{\alpha}(\mi\xi)$ is the ground-state polarisability tensor
and $\tens{G}^{(1)}$ is the scattering Green tensor. We are neglecting
thermal contributions to the CP force, which is a good approximation
for the distances considered in this work. For the structure in Fig.~2
in the paper, the latter can be given in a Rayleigh decomposition
\begin{multline}
\label{Greentensor} \tens{G}^{(1)}(\vec{r},\vec{r}',\omega)
=\frac{\mi}{8\pi^2}\int_{-\pi/d}^{\pi/d}\!\!\dif k_x\!\!
 \sum_{m,n=-\infty}^\infty\int_{-\infty}^\infty\!\!\dif k_y\!\!
\\ \times \sum_{\sigma,\sigma'=E,H}
\frac{\me^{\mi (k_x^mx-k_x^nx')+\mi k_y(y-y')%
+\mi(k_z^{m+}z+k_z^{n-}z)}}{k_z^{n-}} \:
\vec{e}_{m+}^\sigma(k_x,k_y,\omega)
 R_{mn}^{\sigma\sigma'}(k_x,k_y,\omega)
 \vec{e}_{n-}^{\sigma'}(k_x,k_y,\omega)\;.
\end{multline}
Here, $\vec{k}_\pm=(k_x^m,k_y,k_z^{m\pm})$ is the wave vector for
upward/downward moving waves; its $x$- and $z$-components read $k_x^m=k_x+mq$
($q=2\pi/d$: lattice vector of the grating) and
$k_z^{m\pm}=\sqrt{\omega^2/c^2-(k_x^m)^2-k_y^2}$ with
$\operatorname{Im}\,k_z^{m\pm}\gtrless0$; and the polarisation unit vectors are
defined by
\begin{gather}
\label{polarisationvectorE}
\vec{e}_{m\pm}^E(k_x,k_y,\omega)
 =\frac{c}{\omega\sqrt{\omega^2/c^2-k_y^2}}
 \begin{pmatrix}k_x^mk_y\\
 k_y^2-\omega^2/c^2\\ \pm k_yk_z^m
 \end{pmatrix}\;,\\
\label{polarisationvectorH}
\vec{e}_{m\pm}^H(k_x,k_y,\omega)
 =\frac{1}{\sqrt{\omega^2/c^2-k_y^2}}
 \begin{pmatrix}\mp k_z^{m\pm}\\ 0\\ k_x^m\end{pmatrix}\;.
\end{gather}
The latter are chosen such that the $y$-components of the
electric/magnetic fields vanish for $\sigma=H,E$. The Rayleigh
reflection coefficients $R_{mn}^{\sigma\sigma'}(k_x,k_y,\omega)$ are
calculated by numerically integrating the Maxwell equations within
the grating ($0<z<h$) and imposing conditions of continuity at its
upper (free-space) and lower (sapphire) boundaries
\cite{Marachevsky08}. They are even functions of $k_y$ and obey the
following symmetries: $R_{mn}^{\sigma\sigma'}(k_x,k_y,\mi\xi)=\pm
R_{-m-n}^{\sigma\sigma'\ast}(-k_x,k_y,\mi\xi)$ and
$R_{mn}^{\sigma\sigma'}(k_x,k_y,\mi\xi)k_z^{n-}=\pm
R_{-n-m}^{\sigma'\sigma\ast}(k_x,k_y,\mi\xi)k_z^{m+}$ with $+$ for
$\sigma\sigma'=EE,HH$ and $-$ otherwise. Our theory is able to
allow for anisotropic atoms or molecules. For a sufficiently
anisotropic molecule, we find a repulsive CP potential along the
z-axis similar to the repulsive Casimir force predicted in
Ref.~\cite{Levin10} (see also \cite{Milton5}). Note that for the
isotropic atoms used in the current experiment,
$\bm{\alpha}=\alpha\tens{I}$ ($\tens{I}$: unit tensor),
Eq.~(\ref{CPpotential2}) simplifies to Eq.~(1) of the main manuscript
and our formalism reduces to that of Ref.~\cite{Reyes10}.

The evanescent-wave potential is generated by an incoming wave of
(free-space) wavelength $\lambda\!=\!765\,\mathrm{nm}$ which
impinges on the sapphire--grating interface at an incidence angle
$\theta=35.50^\circ=0.6196\,\mathrm{rad}$, with the plane of
incidence being parallel to the grating bars. The components of the
wave vector in the sapphire layer are hence $k_x^m=0$,
$k_y=k\sin\theta=8.39\times 10^6\,\mathrm{m}^{-1}$,
$k_z^{0+}=k\cos\theta=1.18\times 10^7\,\mathrm{m}^{-1}$
($k=\operatorname{Re}n_\mathrm{sapp}\omega/c$ with
$\operatorname{Re}n_\mathrm{sapp}=1.76$ and $\omega=2\pi
c/\lambda=2.46\times10^{15}\mathrm{rad}/\mathrm{s}$). The incoming
field is polarised such that its component perpendicular to the
grating vanishes; it can hence be written as
$\vec{E}_\mathrm{in}(\vec{r})=E_{\mathrm{sapp}}\vec{e}_{0+}^E(0,k_y,\omega)\me^{
\mi(k_yy+k_z^{0+}z)}$. The field amplitude inside sapphire can be
related to the respective laser power $P_{\mathrm{sapp}}$ and beam
waist $w_{\mathrm{sapp}}$ via
$\frac{1}{2}\varepsilon_0n_\mathrm{sapp}cE_\mathrm{sapp}^2=I_\mathrm{sapp}=
P_\mathrm{sapp}/(2\pi w_\mathrm{sapp}^2$). The laser power is in
turn related to its free-space value by
$P_\mathrm{sapp}=T_{\mathrm{free-space}\rightarrow\mathrm{sapp}}
P_\mathrm{free-space}$ where
$T_{\mathrm{free-space}\rightarrow\mathrm{sapp}}^2=0.78$ has been
determined experimentally from a set-up with symmetric light paths.
The measured beam waist of $170\,\mu\mathrm{m}$ in free space
results in an effective beam waist
$w_\mathrm{sapp}=182.9\,\mu\mathrm{m}$ in sapphire after transitions
through the free-space--glass and glass--sapphire interfaces. After
transmission through the grating, the external laser leads to a
field
\begin{equation}
\label{Eevanescent}
\vec{E}(\vec{r})
=E_\mathrm{sapp}
\sum_{n=-\infty}^\infty
 \sum_{\sigma=E,H}\me^{\mi (nqx+k_yy)-\kappa_z^{n+}z}
 \vec{e}_{n+}^{\sigma}(nq,k_y,\omega)
 T_{n0}^{\sigma E}(0,k_y,\omega)
\end{equation}
with $\kappa_z^{n+}=\sqrt{k_y^2-\omega^2/c^2+n^2q^2}\,$ where the Rayleigh
transmission coefficients $T_{mn}^{\sigma\sigma'}(k_x,k_y,\omega)$ are found
from numerical integration. When interacting with a Rb atom, this evanescent
field generates an optical potential
\begin{multline}
\label{Uevanescent}
U_\mathrm{EW}(\vec{r})
=\sum_{i=1,2}\frac{|\vec{d}_i|^2|\vec{E}(\vec{r})|^2}
 {3\hbar\Delta_i}
= \sum_{i=1,2}\frac{|\vec{d}_i|^2E_\mathrm{sapp}^2}{3\hbar\Delta_i}
 \sum_{m,n=-\infty}^\infty
 \sum_{\sigma,\sigma'=E,H}
 \me^{\mi(m-n)qx-(\kappa_z^{m+}+\kappa_z^{n+^*})z} \\
\times \vec{e}_{m+}^{\sigma}(mq,k_y,\omega)\sprod
 \vec{e}_{n+}^{\sigma'^*}(nq,k_y,\omega)
 \: T_{m0}^{\sigma E}(0,k_y,\omega)
 T_{n0}^{\sigma'E^*}(0,k_y,\omega)\;,
\end{multline}
$\omega_1=2.37\times 10^{15}\mathrm{rad}/\mathrm{s}$,
$\omega_2=2.41\times 10^{15}\mathrm{rad}/\mathrm{s}$,
$d_1=2.54\times 10^{-29}\mathrm{Cm}$ and
$d_2=3.58\times 10^{-29}\mathrm{Cm}$ are
the transition frequencies and dipole-matrix elements for the $D_1$
($5^2S_{1/2}\rightarrow 5^2P_{1/2}$) and $D_2$ ($5^2S_{1/2}\rightarrow
5^2P_{3/2}$) lines of the isotropic Rb atom \cite{Steck09}.

 A numerical error
 in the evaluation of the potential is 0.1\%.
 The number of Rayleigh coefficients
 used was selected to obtain the needed accuracy,
 N=30 (2N+1 terms in every Rayleigh expansion)
 is sufficient to obtain the potential with needed accuracy
 even at  closest separations from the grating shown on Fig.2.

\section{Casimir-Polder potential using the Hamaker approach}
The scattering Green tensor $\tens{G}^{(1)} (\vec{r},\vec{r},\omega)$ in
Eq.~(\ref{CPpotential2}) can be expanded in a Born series with respect to a
known reference Green tensor $\tens{G}^{(0)} (\vec{r},\vec{r},\omega)$ as
\cite{Scheel09}
\begin{eqnarray}
\lefteqn{ \tens{G}^{(1)} (\vec{r},\vec{r},\omega)= \frac{\omega^2}{c^2} \int
\mathrm d^3s \, \tens{G}^{(0)} (\vec{r},\vec{s},\omega)
\delta\epsilon(\omega) \tens{G}^{(0)} (\vec{s},\vec{r},\omega)} \nonumber \\
&& + \left(\frac{\omega^2}{c^2}\right)^2 \int \mathrm d^3s \, \mathrm d^3s' \,
\tens{G}^{(0)} (\vec{r},\vec{s},\omega) \delta\epsilon(\omega) \tens{G}^{(0)}
(\vec{s},\vec{s'},\omega) \delta\epsilon(\omega) \tens{G}^{(0)}
(\vec{s'},\vec{r},\omega)+ \dots
\label{eqn:Born}
\end{eqnarray}
where the integration range covers the volume of the material under
investigation. The perturbation $\delta\epsilon(\omega)=\epsilon^{(1)}(\omega)
-\epsilon^{(0)}(\omega)$ denotes the deviation from the reference permittivity.
In the case of the free-space Green tensor as our reference, the
difference permittivity simply equals the susceptibility [here:
$\delta\epsilon(\omega)= \chi_\text{Au}(\omega)$] of the material. The
free-space Green tensor can be written as \cite{Scheel09}
\begin{equation}
\tens{G}^{(0)}(\vec{\varrho},\omega) =
-\frac{c^2}{3\omega^2}\boldsymbol{\delta}(\vec{\varrho}) +\frac{\omega}{4 \pi c}
\left[ f \left( \frac{c}{\omega \varrho} \right) \tens{I} - g \left(
\frac{c}{\omega \varrho}\right) \frac{\vec{\varrho} \otimes
\vec{\varrho}}{\varrho^2} \right] \me^{ \mi \varrho \omega/c}
\end{equation}
with $f(x) = x+ \mi x^2 - x^3$ and $g(x) = x+ 3 \mi x^2 -3 x^3$.

The Hamaker approach only uses the first term of the Born series expansion.
Because the Born series is a perturbative expansion in the susceptibility
$\chi(\omega)$, it converges badly for materials with a large susceptibility
such as gold. To improve this convergence, and to implement a local-field
correction, the reference Green tensor is separated into a
regular $\tens{R}$ and a singular part
$\tens{D}=-\frac{c^2}{3\omega^2}\boldsymbol{\delta}(\vec{\varrho})$. Inserted
into Eq.~(\ref{eqn:Born}) and performing the integrals over the
$\delta$-distributions yields the local-field corrected first-order scattering
Green tensor
\begin{equation}
\tens{G}^{(1)} (\vec{r},\vec{r},\omega) = \frac{\omega^2}{c^2} \chi(\omega)
\sum_{n=0}^\infty \left(-\frac{1}{3}\chi(\omega)\right)^n \int \mathrm d^3s
\,\tens{R}^{(0)} (\vec{r},\vec{s},\omega) \tens{R}^{(0)}
(\vec{s},\vec{r},\omega) \,,
\label{eqn:Born2}
\end{equation}
which leads to
\begin{equation}
\label{eq:Born3}
\tens{G}^{(1)}(\vec{\varrho},\omega)=\frac{\omega^2}{c^2}
\frac{\chi(\omega)}{1+\chi(\omega)/3} \int \mathrm d^3s\, \tens{R}^{(0)}
(\vec{r},\vec{s},\omega) \tens{R}^{(0)} (\vec{s},\vec{r},\omega)
\end{equation}
with
\begin{equation}
\tens{R}^{(0)} (\vec{\varrho},\omega)=\frac{\omega}{4 \pi c} \left[ f
\left( \frac{c}{\omega \varrho} \right) \tens{I} - g \left( \frac{c}{\omega
\varrho}\right) \frac{\vec{\varrho} \otimes \vec{\varrho}}{\varrho^2} \right]
\me^{ \mi \varrho \omega/c}\,.
\end{equation}
Eq.~(\ref{eq:Born3}) has to be evaluated numerically. Because of the smooth
shape of the potential, numerical integration methods converge quickly with
errors proportional to at most the curvature of the potential, which is small in
the range of investigation.

In the Hamaker approach (pairwise summation) the potential is clearly additive.
The total potential can be separated into three parts: one related to the
evanescent field, one related to the dispersion interaction between Rb and Au,
and one part for the Casimir-Polder interaction between the Rb atoms and the
sapphire substrate. The latter can be neglected, because of the vastly
different susceptibilities of gold and sapphire.

\section{BEC preparation}
The BEC is prepared in an ultra-high vacuum chamber. After precooling and
trapping $^{87}$Rb atoms in a magneto-optic trap the atoms are adiabatically
transferred into a highly compressed magnetic trap, where they are further
cooled by forced radio-frequency evaporation by which Bose-Einstein condensates
with typically $2\times 10^5$ atoms are generated. This is done at a distance of
several hundred $\mu\textrm{m}$ from the surface of a dieletric glass prism. A
thin sapphire substrate containing the gold grating investigated in this paper
is glued to the top of this prism. After preparation of the BEC we suddenly
displace the magnetic trapping minimum by a distance $\Delta z$ raising thereby
the potential energy of the BEC by $\Delta E=\frac{1}{2}m\omega\Delta z^2$ with
atomic mass $m$ and magnetic trapping frequency $\omega$. The BEC is accelerated
towards the new trapping minimum. After a quarter of the oscillation period
$\delta t=\frac{\pi}{2\omega}$ we switch-off the magnetic trap and apply a
constant magnetic field gradient of $B^\prime=15~\textrm{Gcm}^{-1}$ which
compensates the gravitational force. Thus, the atoms are not further accelerated
or decelerated due to gravitation while moving towards the grating. We measure
the actual velocity of the atoms by taking absorption images of the cloud at
several time intervals of $1~\textrm{ms}$ after the acceleration.